\documentclass[%
 aip,%
 amsmath,amssymb,
reprint,%
apl,
twocolumn,
superscriptaddress
]{revtex4}

\usepackage{graphicx}
\usepackage{bm}

\begin{document}

\title{Interface effects on an ultrathin Co film in multilayers based on the organic semiconductor Alq$_3$}

\author{A. A. Sidorenko}
 \altaffiliation[Present address: ]{Institute of Solid State Physics, TU Wien, A - 1040 Vienna, Austria}
\email{sidorenko@ifp.tuwien.ac.at}
\author{C. Pernechele}
\author{P. Lupo}
\author{M. Ghidini}
\author{M. Solzi}
\author{R. \surname{De Renzi}}
\affiliation{Dipartimento di Fisica, Universit\`a di Parma, I-43124 Parma, Italy}

\author{I. Bergenti}
\author{P. Graziosi}
\author{V. Dediu}
\affiliation{ISMN-CNR, I-40129 Bologna, Italy}

\author{L. Hueso}
\affiliation{CIC nanoGUNE Consolider, E-20018 Donostia San Sebastian, Spain}
\affiliation{IKERBASQUE, Basque Foundation for Science, E-48011 Bilbao, Spain}

\author{A.T. Hindmarch}
\affiliation{School of Physics and Astronomy, University of Leeds, Leeds, LS2 9JT, United Kingdom}

\date{\today}

\begin{abstract}
The effect of the AlO$_x$ barrier thickness on magnetic and morphological properties of Ta/Co/(AlO$_x$)/Alq$_3$/Si hybrid structures was systematically studied by means of atomic force microscopy,  SQUID magnetometry and nuclear magnetic resonance (NMR). All used techniques pointed out that the barrier thickness of 2 nm is required to obtain a magnetically good cobalt layer on top of Alq$_3$. $^{59}$Co NMR measurements revealed that the AlO$_x$ barrier gives rise to the formation of an interface layer with ``defective'' cobalt favouring growth of ``bulk'' cobalt with good magnetic properties. 
\end{abstract}

\pacs{76.60.-k, 75.70.-i, 75.30.Gw}

\maketitle

Spin injection and transport in organic semiconductors (OS) have been attracting much attention since the demonstration of the magnetoresistive effects in lateral devices\cite{Dediu2002181} and the spin valve (SV) effect in the vertical configuration\cite{Xiong2004}. These hybrid structures provide tunable electronic properties combined with low-cost production, efficient spin-injection and long spin diffusion length.\cite{Dediu2009} A crucial point to take into account in the case of vertical architecture of SVs is the quality of the top ferromagnetic (FM) electrode that is deposited on the OS layer. As expected, the magnetic properties and their reproducibility are strongly affected by the underlayer roughness.\cite{Bergenti2007,Pernechele2009,Liu2009} Furthermore, it has been shown that deposited Co, both as clusters or atoms, can penetrate into the Alq$_3$, causing pinholes and Co inclusions over a distance of $\sim100$ nm.\cite{Xiong2004} On the other hand, transmission electron microscopy revealed an abrupt interface between Co and Alq$_3$ with minimal intermixing.\cite{Xu2009} Therefore the insertion of a buffer layer between the FM top electrode and the OS has been widely adopted to reduce possible interface disorder.\cite{dediu2008} Following this approach, examples of efficient device performance are reported in Ref.~\onlinecite{dediu2008} where the good quality of the Alq$_3$ layer without Co inclusions and with significantly better spin-valve effect was demonstrated in Co/tunnel barrier/Alq$_3$/La$_{0.67}$Sr$_{0.33}$MnO$_3$(LSMO) structure after insertion of a thin Al$_2$O$_3$ tunnel barrier. Since magnetic and transport properties of FM/OS/FM junctions are strongly dependent on interface roughness or compositional intermixing, a deep comprehension of the interface quality is essential to understand the device performance and operation conditions. 

In this letter we study magnetic and morphological properties of Co/(AlO$_x$)/Alq$_3$ interfaces with various AlO$_x$ barrier thickness probed by means of zero-field $^{59}$Co nuclear magnetic resonance (NMR), superconducting quantum interference device magnetometry (SQUID), and atomic force microscopy (AFM). The interfaces under study represent the top electrode of vertical organic SVs layers of Co/Alq$_3$/LSMO where the insertion of a buffer layer of aluminum oxide was essential for devices operation.

50 nm thick Alq$_3$ films  were deposited at room temperature (RT) by organic molecular beam deposition in UHV conditions (10$^{-9}$ - 10$^{-10}$  mbar) on naturally oxidized Si substrates. The Alq$_3$ layer was covered by AlO$_x$ tunnel barrier (0-5 nm) grown by Channel Spark Ablation from a polycrystalline  stoichiometric target. The 15-nm thick Co top layer was deposited by magnetron sputtering in Ar atmosphere. Finally, a 2-nm thick Ta capping overlayer was added in order to prevent Co oxidation. These systems lack the bottom LSMO electrode in order to perform a magnetic characterization, but 50 nm OS film thickness ensures that the Alq$_3$ surface morphology is independent from the substrate (both Si or LSMO) and the Co/Alq$_3$ interfaces can be compared with the analogous ones in the full operating devices.

\begin{figure}
\includegraphics{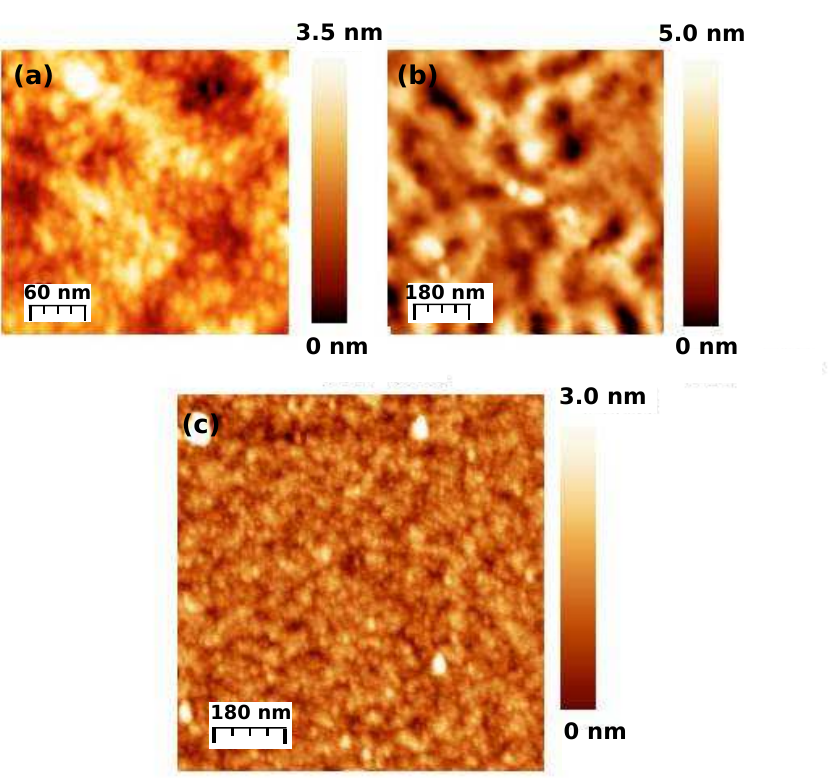}
\caption{\label{fig:AFM_one} AFM Images (height topographic signal) for the sample Ta/Co/Alq$_3$/Si (top panel):  a) example of a typical Ta/Co morphology (zoom from the original size 5x5 $\mu$m); b) example of Alq$_3$ recrystallization when the Co barely decorates the Alq$_3$ structure underneath (zoom from the original size 5x5 $\mu$m), and c) Ta/Co/1nm AlO$_x$/Alq$_3$/ Si (bottom panel).}
\end{figure}

The sample Co/Alq$_3$/LSMO without AlO$_x$ barrier demonstrates a quite inhomogeneous morphology with different surface microstructures examined by AFM (Nanoscope III/A Multimode) in tapping mode. Some part of the sample has a nanograined morphology resembling the typical microstructure of cobalt deposited on ordinary substrates, e.g. Si  (see Fig.~\ref{fig:AFM_one}a). The other part of the sample, however, bears clear resemblance to the morphology of recrystallized Alq$_3$ (see Fig.~\ref{fig:AFM_one}b). We note that the bare Alq$_3$ thin films is amorphous with  root-mean-square (rms) roughness of about 1.5 nm (data not shown), but exposure to air is known to induce the formation of Alq$_3$ crystalline structures increasing thus the overall film roughness.\cite{Aziz1998} The estimated rms roughness of Co/Alq$_3$/LSMO is higher in zones where Co decorates Alq$_3$ with peak-to-peak (PP) of about 10 nm, as compared with the continuous Co regions where PP is less or equal to 2 nm. This large contrast confirms  the morphological indication that Co adhesion depends on the crystallinity of Alq$_3$. The absolute value of the roughness may however be influenced by Ta capping layer, since tantalum atoms tend to fill the small gaps between large cobalt particles, smoothing the measured  surface.\cite{clavero2008} The average grain size in the smoother areas is $30-40$ nm.

\begin{figure}
\includegraphics{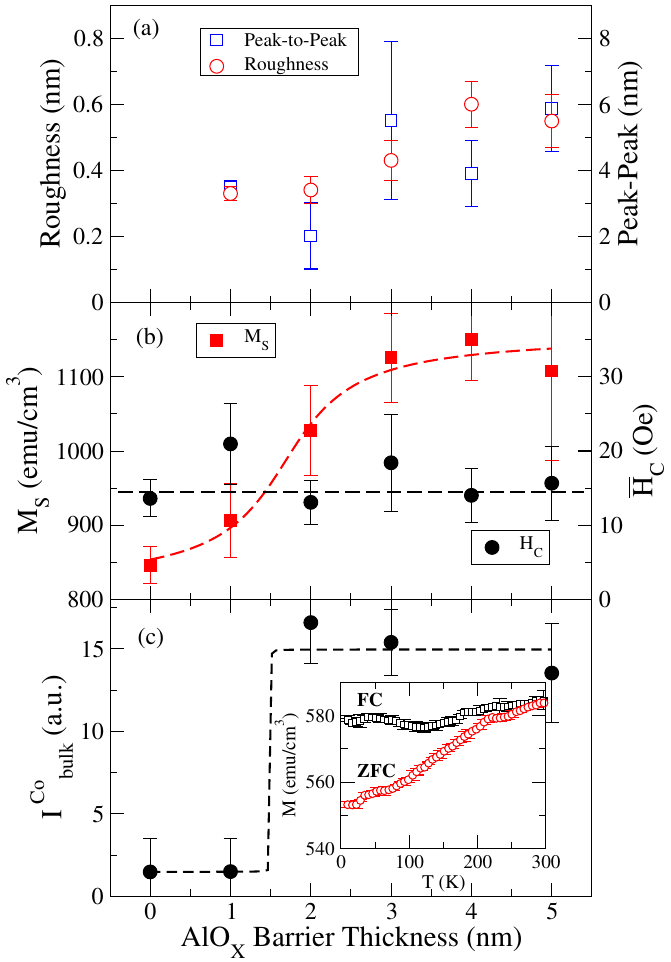}
\caption{\label{fig:Barrier} The AlO$_x$ barrier thickness dependence of a) rms roughness (circles) and PP height (squares); b) saturation magnetization M$_S$ (squares) and coercive field H$_C$ (dots); c) $^{59}$Co NMR integrated intensity of 'bulk' cobalt (dots). Lines are guides for eyes. Inset: temperature dependence of the magnetization measured in the sample without barrier (H=10 Oe, in-plane).}
\end{figure}

Inserting an AlO$_x$ barrier makes the morphology of the Co film and Ta capping layer fairly homogeneous (see Fig.~\ref{fig:AFM_one}c). In the sample with 1-nm thick AlO$_x$ barrier the surface is
smooth with an rms roughness of $0.33\pm0.02$ nm, PP height of $3.5\pm0.2$ nm, and an average grain size $\sim$30 nm. However further thickening of the AlO$_x$ spacer does not change the surface morphology of the samples significantly. Fig.~\ref{fig:Barrier}(a) displays the rms roughness and PP variation of the topographic height versus thickness of the AlO$_x$ barrier, in the range $1-5$ nm. Both parameters are nearly constant albeit for a small increase with thickness as expected for an overall thicker film. These AFM measurements draw us to the conclusion that the AlO$_{x}$ barrier of 1 nm is already sufficient to guarantee a smooth and morphologically homogeneous Co film. 

The AlO$_x$ barrier improves also the magnetization of the heterostructures. All samples, measured by a Quantum Design MPMS-XL 5T SQUID, reproduce a FM behaviour at RT (data not shown). The saturation magnetization from $M_S\approx850\pm20$ emu/cm$^3$ for the sample without barrier increases with the AlO$_x$ thickness (Fig.~\ref{fig:Barrier}(b)) recovering the best value $M_S\approx1100$ emu/cm$^3$ obtained by deposition of Co on Si substrate. This recovery is achieved for barrier thickness of 2 nm or more determining a critical thickness for AlO$_x$ barrier required to obtain Co film of a good quality. This value is consistent with the threshold for a continuous Co film obtained from AFM. The low coercive field value, $\overline{H}_C\approx15$ Oe, suggests the nanocrystalline nature of these films with ferromagnetically weakly coupled Co nanograins and strongly reduced anisotropy,\cite{Herzer} and the absence of variation with AlO$_x$ thickness nicely correlates  with the almost constant roughness of Fig.~\ref{fig:Barrier}(a).

\begin{figure}
\includegraphics{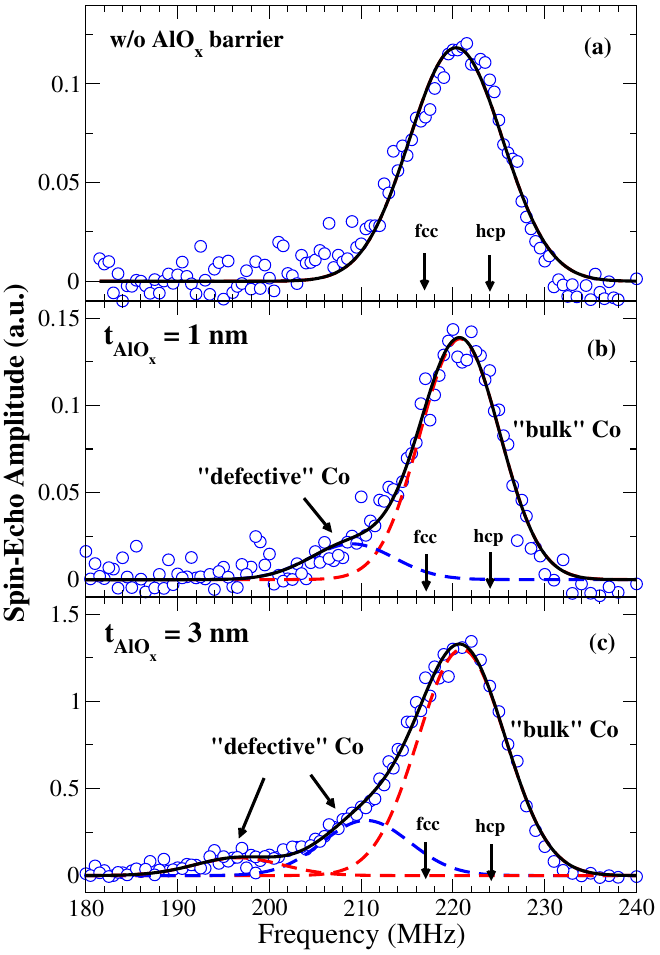}
\caption{\label{fig:NMR_Spec} Zero-field $^{59}$Co NMR in Ta/Co/(AlO$_x$)/Alq$_3$/Si a) without, b) with 1-nm thick, and c) with 3-nm thick AlO$_x$ barrier measured at T=1.6 K.  All spectra were corrected by corresponding enhancement factor.}
\end{figure}

A threshold barrier thickness of $\sim2$ nm is also found in zero-field $^{59}$Co NMR measured at T=1.6 K with home-built broad-band NMR spectrometer\cite{Allodi} and a tuned probe circuit. All spectra (see Fig.~\ref{fig:NMR_Spec}), obtained on the same set of samples, show the presence of a main Gaussian resonance line centered in the usual Co metal range, at a frequency of $\sim$220 MHz. 
The full width at half maximum is $\sim$10 MHz, i.e. the line is much broader than in perfect, epitaxial films, but it can still be attributed exclusively to ``bulk'' Co nuclei, surrounded by the full set of z=12 nearest neighbour (nn) Co.\cite{Panissod1998} The width is due to the presence of smaller perturbations, such as a mixture of  \textit{fcc} Co ($\sim$217.4 MHz), \textit{hcp} Co ($\sim$224 MHz), and a large amount of stacking faults (distributed between \textit{fcc} and \textit{hcp} phases).\cite{Panissod1998} However, the samples without and with 1-nm  AlO$_x$ barrier reveal the much lower NMR signal intensity I$^{Co}_{bulk}$  than samples with the thicker barrier (see Fig.~\ref{fig:NMR_Spec} and Fig.~\ref{fig:Barrier}(c)). Since the observed NMR signal arises from domain walls (DW) (as evident by the large detected DW enhancement)\cite{Panissod1998}, the reduction of I$^{Co}_{bulk}$ for barrier thickness below 2 nm may be due to the formation of very small single domain cobalt particles (grains) disconnected magnetically. Therefore, such an effect in the samples with the barrier thinner than $\sim$2 nm may be responsible for the reduced M$_S$ values (compare Fig.~\ref{fig:Barrier}(b) and (c)).

This assumption is supported also by the temperature dependence of the magnetization measured in the junction without barrier at H=10 Oe. The inset of Fig.~\ref{fig:Barrier}(c) shows that for this film zero field cooling (ZFC) and field cooling (FC) M(T) curves, both  measured on heating, split below $\sim220$ K. The observed behaviour could be interpreted as the superposition of a cluster-like contribution and a normal FM one. We suppose that at interface with Alq$_3$ small Co clusters are present, likely interacting both with other clusters and with the continuous film rising above. In that situation a collective behaviour is expected and accordingly the temperature at which divergence between ZFC and FC curves occurs indicates the onset of cooperative freezing of interacting magnetic clusters.\cite{Bedanta2009} It is important to notice that for the samples with the AlO$_x$ barrier the measured temperature dependence of FC/ZFC curves is clearly different.\cite{SupplMat} While we cannot exclude the presence of a few Co clusters even in the case of samples with the oxide barrier, as some voids or defects in the AlO$_x$ layer are likely, we think that these clusters should be larger with respect to the case of bare Alq$_3$, or strongly interacting, as they appear to be blocked up to RT. Hence the presence of a 2-nm thick AlO$_x$ barrier is sufficient to grow magnetically well connected FM Co layer on top of Alq$_3$.

In addition the spectra obtained on samples with AlO$_x$ barrier contain a distinct, structured lower frequency shoulder (see Fig.~\ref{fig:NMR_Spec}) attributed to the ``defective'' cobalt. Such a ``defective'' cobalt have one or more either missing or substituted nn Co (e.g., with Al ion). The replacement of one nn Co atom by a non-magnetic atom or defect removes a hyperfine contribution at the cobalt nucleus producing a shift of the resonance $^{59}$Co frequency by about 12-22 MHz depending on substituent.\cite{Panissod1998} Therefore inserting the AlO$_x$ barrier results in the formation of Co$_{1-c}$X$_c$ alloy (c is the concentration of the defects X) presumably located close to the interface. This observation and the fact that the thick enough AlO$_x$ barrier substantially improves the magnetization and NMR intensity I$^{Co}_{bulk}$ (see Fig.~\ref{fig:Barrier}(b) and (c)) allow to conclude that the formation of an interface layer with ``defective'' Co (Co$_{1-c}$X$_c$ random alloy) is important to obtain a good quality FM ``bulk'' cobalt on top of Alq$_3$ layer. 

This work has been funded by FPVI STREP OFSPIN Grant No. 033370.

\end{document}